\pdfoutput=1
\documentclass[11pt,a4paper]{article}
\usepackage{jheppub}

\usepackage{amssymb}
\usepackage{bm}
\usepackage{subfigure}	
\usepackage{graphicx}
\usepackage{dcolumn}
\usepackage{multirow}
\usepackage{mathptmx}

\usepackage{url}
\usepackage{footnote}

\DeclareGraphicsExtensions{.pdf}

\def\LQ{\mathrm{LQ}}

\def\Zll{\ensuremath{Z/\gamma^{*} \rightarrow \ell\ell}}

\def\Wboson{\ensuremath{W}}
\def\Zboson{\ensuremath{Z}}
\def\ZGamma{\ensuremath{Z/\gamma^{*}}}
\def\ZGammaee{\ensuremath{Z/\gamma^{*} \rightarrow ee}}
\def\ZGammamm{\ensuremath{Z/\gamma^{*} \rightarrow \mu\mu}}
\def\ZGammatau{\ensuremath{Z/\gamma^{*} \rightarrow \tau\tau}}

\def\ZGammallMargar{\ensuremath{Z/\gamma^{*} \rightarrow \ell\ell}}

\def\pt{\ensuremath{p_{\mathrm{T}}}} 
 
\def\et{\ensuremath{E_{\mathrm{T}}}} 
 
\def\ET{\ensuremath{E_{\mathrm{T}}}} 
 
\def\ST{\ensuremath{S_{\mathrm{T}}}} 
\def\ptsq{\ensuremath{p^2_{\mathrm{T}}}} 
 
\def\met{\ensuremath{E_{\mathrm{T}}^{\mathrm{miss}}}} 

\def\TeV{\ifmmode {\mathrm{\ Te\kern -0.1em V}}\else
                   \textrm{Te\kern -0.1em V}\fi}
\def\GeV{\ifmmode {\mathrm{\ Ge\kern -0.1em V}}\else
                   \textrm{Ge\kern -0.1em V}\fi}
\def\MeV{\ifmmode {\mathrm{\ Me\kern -0.1em V}}\else
                   \textrm{Me\kern -0.1em V}\fi}

\def\antibar#1{\ensuremath{#1\bar{#1}}}

\def\ttbar{\antibar{t}}

\def\bbbar{\antibar{b}}

\def\qqbar{\antibar{q}}

\def\ifb{\mbox{fb$^{-1}$}}
\def\mass#1{\ensuremath{m_{#1#1}}}
\def\twomass#1#2{\ensuremath{m_{#1#2}}}
\def\rts {\ensuremath{\sqrt{s}}}

\mathchardef\mhyphen="2D

\def\tauhadvis{\ensuremath{ \tau_{\mathrm{had \mhyphen vis}}}}

\usepackage{preprintcover}  

\PreprintCoverPaperTitle{Search for third generation scalar leptoquarks in $\boldsymbol{pp}$ collisions at $\boldsymbol{\rts}$ = 7 TeV with the ATLAS detector}  

\PreprintIdNumber{CERN-PH-EP-2012-317} 

\PreprintCoverAbstract{A search for pair-produced third generation scalar leptoquarks is presented, using proton--proton collisions at \rts~=~7 \TeV~at the LHC.  The data were recorded with the ATLAS detector and correspond to an integrated luminosity of 4.7\,\ifb.  Each leptoquark is assumed to decay to a tau lepton and a $b$-quark with a branching fraction equal to 100\%.  No statistically significant excess above the Standard Model expectation is observed.  Third generation leptoquarks are therefore excluded at 95\% confidence level for masses less than 534 \GeV.}  

\PreprintJournalName{JHEP}

\title{Search for third generation scalar leptoquarks in $\boldsymbol{pp}$ collisions at $\boldsymbol{\rts}$ = 7 TeV with the ATLAS detector}

\author{The ATLAS Collaboration}

\date{\today}

\abstract{ A search for pair-produced third generation scalar leptoquarks is presented, using proton--proton collisions at \rts~=~7 \TeV~at the LHC.  The data were recorded with the ATLAS detector and correspond to an integrated luminosity of 4.7\,\ifb.  Each leptoquark is assumed to decay to a tau lepton and a $b$-quark with a branching fraction equal to 100\%.  No statistically significant excess above the Standard Model expectation is observed.  Third generation leptoquarks are therefore excluded at 95\% confidence level for masses less than 534 \GeV.}

\begin{document}
\maketitle
\flushbottom

\sloppy

\section{Introduction}
Leptoquarks (LQ) are colour-triplet  bosons that carry both lepton and
baryon  numbers and  have  a fractional  electric  charge. They  are
predicted    by    many    extensions    of   the    Standard    Model
(SM)~\cite{BSM,techni2,techni3,string,comp,pati_salam_colour,georgi_glashow_unification}  and may provide unification between the quark and lepton sectors.  In accordance  with experimental results
on  lepton-number  violation,  flavour-changing neutral  currents  and
proton  decay,  it  is  assumed  that individual leptoquarks  do  not  couple  to
particles  from  different  generations~\cite{mBRW1,mBRW}, thus leading to three generations of leptoquarks.   The  most
recent limit on pair-produced  third generation scalar leptoquarks ($LQ_3$) decaying to $\tau b \tau b$~comes from
the CMS experiment, in which scalar leptoquarks with masses below 525 \GeV~are excluded at the 95\% confidence level~(CL)\cite{CMS_LQ3}.  Limits have also been set by the D0~\cite{D0} and CDF~\cite{CDF_Scalar_top} experiments at the Tevatron, which have excluded third generation scalar leptoquarks with masses up to 210 \GeV~and 153~\GeV~respectively.  First and second generation scalar leptoquarks have been excluded up to 830 \GeV~and 840 \GeV~respectively \cite{CMS_LQ1-2,LQ1,LQ2}.  The  results  presented  here  are  based  on  a  total
integrated luminosity of 4.7\,\ifb~of proton-proton collision data at
a  centre-of-mass  energy of \rts~=~7\,\TeV, collected  by  the  ATLAS
detector at the LHC during 2011.  The final states investigated arise from
the decay of both leptoquarks into a tau lepton and a $b$-quark, leading to
a $\tau b \tau b$ final state.  The branching fraction of $LQ_{3}$ decays to 
$\tau b$ is assumed to be  equal to 100\%. 

Tau leptons can decay either leptonically (to an electron or muon plus two neutrinos), or hadronically (typically to one or three charged hadrons, plus one neutrino, and zero to four neutral hadrons).  Since the final state includes two taus, this leads to three possible sub-categories of events: di-lepton, lepton--hadron and hadron--hadron.  Of these, the  lepton--hadron category  has  the  largest
branching fraction (45.6\%), and the presence of one  charged light  lepton ($\ell=e,\mu$)  in  the event is useful for  event  triggering and provides better
rejection of  the multi-jets background.  Only the lepton--hadron decay mode is considered in this analysis,
resulting  in  either an $e   \tauhadvis  bb+3\nu$  or  $\mu  \tauhadvis
bb+3\nu$ final state, where \tauhadvis~refers to the visible (non-neutrino) components of the hadronic tau decay.  
 
Selected  events  are  therefore required to have one electron or muon with large transverse momentum~(\pt),
one high-\pt~hadronically decaying tau, missing transverse energy from the
tau   decays,  and   two  high-\pt~jets.  Searches   are  performed
independently  for the electron  and muon  channels.  The  results are
subsequently combined and interpreted as lower bounds on the $LQ_{3}$ mass.

\section{The ATLAS detector}
\label{sec:detector}
The  ATLAS detector~\cite{det2}  is  a multi-purpose  detector with  a
forward-backward  symmetric  cylindrical  geometry and  nearly  4$\pi$
coverage in solid  angle.  ATLAS uses a right-handed coordinate
system with its origin at the nominal interaction point (IP) and the
$z$-axis along the beam pipe. The $x$-axis points from the IP to the
centre of the LHC ring, and the $y$-axis points upward.  Cylindrical
coordinates~($r,\phi$) are used in the transverse $(x,y)$~plane, with $\phi$
being the azimuthal angle around the beam pipe.  The pseudorapidity $\eta$
is defined in  terms of the polar angle $\theta$ by  $\rm \eta = -ln
(tan(\theta/2))$.  The three major sub-components of ATLAS are the
tracking detector, the calorimeter and the muon spectrometer.  

Charged particle  tracks and  vertices  are reconstructed  using silicon pixel and microstrip detectors  covering the range  $|\eta|$~$<$~2.5,  and by a straw tube tracker that covers $|\eta|$~$<$~2.0. Electron identification capability is added by employing Xenon gas to detect transition radiation photons created in a radiator between the straws. The inner tracking system is immersed in a  homogeneous 2\,T magnetic field provided by a solenoid.  

Electron,  jet  and   tau  energies  are  measured  in  the calorimeter.  The ATLAS calorimeter system covers a pseudorapidity range of $|\eta|$~$<$~4.9.  Within the region  $|\eta|$~$<$~3.2, electromagnetic calorimetry is provided by barrel and end-cap high-granularity lead/liquid argon (LAr) electromagnetic (EM) calorimeters, with an additional thin LAr presampler covering $|\eta|$~$<$~1.8, to correct for energy loss in material upstream of the calorimeters. Hadronic calorimetry is provided by a steel/scintillator-tile calorimeter, segmented into three barrel structures within $|\eta|$~$<$~1.7, and two copper/LAr hadronic end-cap calorimeters. The forward region (3.1~$<$~$|\eta|$~$<$~4.9) is instrumented by a LAr calorimeter with copper (EM) and tungsten (hadronic) absorbers.

Surrounding  the  calorimeters,  a  muon spectrometer  with  air-core toroids, a  system of precision tracking chambers providing coverage over $|\eta|$~$<$~2.7,  and detectors with triggering  capabilities over $|\eta|$~$<$~2.4 to provide   precise  muon  identification  and momentum measurements. 

A  three-level event-triggering system selects inclusive electron and muon candidates to be recorded for offline analysis.

\section{Monte Carlo simulations}
\label{sec:MC}
Simulated signal  events  are  produced  using the  PYTHIA 6.425~\cite{PYTHIA}  event generator with  underlying-event Tune D6~\cite{D6}  and  CTEQ6L1~\cite{CTEQ}  parton distribution  functions (PDFs).   The  coupling $\lambda_{LQ \rightarrow lq}$ which determines the LQ lifetime and width \cite{LQlq_coupling} is set  to 0.01$\times$4$\pi  \alpha$, where  $\alpha$  is the fine-structure  constant. This  value  is widely  used in  leptoquark searches  and  gives  the leptoquark  a  full width  of  less  than  1\,\MeV~and a decay length of less than 1 mm.  The signal process is normalised using next-to-leading-order  (NLO)  cross-sections  for scalar leptoquark  pair production~\cite{LQ_NLO_XSec}.  The signal production cross-section for a leptoquark mass of 500\,\GeV~is 46.2 fb.

Background processes considered in  this analysis are the production of \Wboson+jets, \ZGamma+jets, \ttbar, single top quarks, boson pairs, and  multi-jets. The  \Wboson- and \Zboson-boson  processes are simulated using the ALPGEN 2.13 generator~\cite{Alpgen} with the technique described in ref.~\cite{MLM} to match the hard process (calculated with a leading-order (LO) matrix element for up to five partons)  to the parton shower of HERWIG 6.510~\cite{Herwig}, and uses JIMMY 4.31~\cite{JIMMY} to model the underlying event.  Wherever available, dedicated ALPGEN 2.13 samples with massive charm and bottom partons were used for the \Wboson+jets and \ZGamma+jets processes.  All samples listed above  are generated  using  the CTEQ6L1 PDFs.  Di-boson  processes ($WW$, $WZ$, and $ZZ$) are modelled with HERWIG 6.510  using the MRST~\cite{MRST} LO PDFs.   Samples of top-quark pair production and  associated production  of  single top quark ($Wt$)  events  are  produced  using  the MC@NLO 4.01~\cite{MCatNLO_4_0,MCatNLO,MCatNLO_tt,MCatNLO_Wt}  generator  interfaced  with  HERWIG 6.510 for  parton showering, and JIMMY 4.31 to  model the underlying event.  The CT10~\cite{CT10} PDFs are used.   Single-top $s$-  and $t$-channel processes are modelled with AcerMC 3.8~\cite{acer} using the MRST LO PDFs.  The top-quark mass is taken as 172.5\,\GeV.  In all  simulated samples  TAUOLA~\cite{tauola}  and PHOTOS~\cite{photos} are used to model $\tau$-lepton decays and additional photon radiation from  charged leptons, respectively.   The \Wboson+jets and \Zboson+jets samples are normalised to the inclusive NNLO cross-sections in the proportions predicted by NLO calculations for exclusive $n$-parton production.  The most precise available calculation (nearly NNLO) is used to normalise \ttbar~production~\cite{ttbar_xSec}.  All other sources of background are normalised using the cross-sections calculated at NLO.

Signal  and  background  events  are  processed  through  a  detailed detector simulation~\cite{ATLASSim} based on GEANT4~\cite{Geant4}. The data  used in  this paper  are  affected by  multiple $pp$  collisions occurring  in the same  or neighbouring  bunch crossings  (pile-up) and have an  average of ten interactions per bunch  crossing.  The effects of pile-up are taken  into account by overlaying simulated minimum-bias events onto the simulated hard-scattering events.  The Monte Carlo (MC) samples are  then   re-weighted  such  that  the  average   number  of  pile-up interactions matches that seen in the data.

\section{Physics object identification}
\label{sec:objID}
Collision  events are  required  to have  at  least one  reconstructed vertex with at least four associated tracks with \pt~$>$~0.4\,\GeV. In events where more than one  vertex is found, the primary  vertex is defined as the one with the highest $\sum \ptsq$ of the associated tracks.  For the final state of interest described below, this choice of primary vertex is correct in 98.9\% (98.4\%) of the cases in the electron (muon) channel for the luminosity range considered here.

Electron candidates  are reconstructed from  clusters of cells  in the electromagnetic  calorimeter and  from tracks in the  inner detector.  They are required to pass a set of electron identification cuts, based on information about the transverse shower shape, the  longitudinal leakage into  the hadronic calorimeter, transition radiation, and the requirement that a good-quality track with a hit in the innermost pixel layer points to the calorimeter cluster~\cite{elID}.  A tight working point corresponding to a selection efficiency of approximately 80\% for true electrons in simulation is chosen.  Electrons are required to have \pt~$>$~20\,\GeV~and $|\eta|$~$<$~2.47, excluding the transition region between the barrel and the end-cap calorimeters, \emph{i.e.}~1.37~$<$~$|\eta|$~$<$~1.52.  Isolation  requirements are  placed on the electron candidates by demanding that the calorimeter transverse energy in a cone of radius $\Delta R$~=~0.2 around the electron (not including  the electron cluster) must be less than  20\% of the electron \pt, where $\Delta R$~=~$\sqrt{(\Delta \eta)^2+(\Delta \phi)^2}$.  In addition, track isolation is imposed by requiring that the \pt~sum of additional tracks (not including the electron track) in a cone of radius $\Delta R$~=~0.2 is less than 20\% of the electron track \pt.  

Muon tracks are reconstructed  independently in the inner detector and in the muon spectrometer.  Tracks are required to have a minimum number of hits in each, and must be compatible in terms of geometrical and momentum matching.  The information from both systems is then used in a combined fit to refine the measurement of the momentum of each muon.  Muon candidates are required to  have \pt~$>$~20\,\GeV~and $|\eta|$~$<$~2.5.  The average muon reconstruction efficiency is approximately 90\%, except for small regions in pseudorapidity where it drops to 80\% \cite{mueff}.  Isolation requirements are imposed  by demanding that the transverse energy (\et) deposited  in  the  calorimeter  in  a  cone  of  radius $\Delta R$~=~0.2 around  the muon (not  including the cells crossed by the  muon) is less than 20\% of the muon \pt.  Furthermore, track isolation  is imposed  by requiring  that the \pt~sum of additional tracks (not including the muon track) in a cone of radius $\Delta R$~=~0.2 be less than 20\% of the muon \pt.

Jets  are  reconstructed  using the  anti-$k_t$~\cite{AntiKt} algorithm with a radius parameter $R$~=~0.4.  The jet algorithm is run on calibrated  topological clusters of  calorimeter cells~\cite{LArClusters}.  Additional \pt-  and $\eta$-dependent corrections  are applied  to jets  to bring them  to  the final  calibrated energy scale~\cite{JES}.   Selected  jets must have \pt~$>$~25\,\GeV~and  $|\eta|$~$<$~2.8. 

The identification  of jets  originating from $b$-quarks  is performed using  a  neural-network-based  tagger~\cite{MV1} that uses the output weights  of   several  likelihood-based  algorithms  as   inputs.   The  track transverse  and longitudinal  impact  parameters with  respect to  the primary vertex are examples of  variables used by these algorithms.  A working point corresponding to an identification efficiency of approximately 70\% for true $b$-jets in simulation is chosen. For jets initiated by gluons or light quarks, the rejection factor (1/fraction that pass the $b$-tagging ID) is of order 100.

The reconstruction  of hadronically decaying tau leptons is seeded by jets  which  are reconstructed  within  the  acceptance  of the  inner detector.  Only  clusters in a cone with  radius $\Delta R$~$=$~0.2 are used to define  the visible tau energy and direction because the products of hadronic tau decays are more collimated than those from multi-jet processes.   Additional corrections depending on the \pt, $\eta$ and  number of tracks are  applied to  bring the reconstructed tau candidates to the correct energy scale \cite{tauID}.  The energy deposition in the  calorimeter is required to  be matched to either  one or three tracks in  the inner  detector.  Hadronically decaying taus are required to have visible \pt~$>$~20\,\GeV, $|\eta|$~$<$~2.5 and unit charge, and are identified using  a   Boosted   Decision   Tree (BDT)\footnote{A BDT is a multivariate analysis technique where the selection is based on a majority vote on the result of several decision trees, each of which is derived from the same training sample by supplying different event weights during the training.}~\cite{BDT}  which  uses   both  calorimeter  and  tracking-based variables  such as  shower width  and track  multiplicity.   A working point with an identification efficiency for true hadronic tau decays in simulation of $\sim 50\%$ is chosen.   The rejection factor for  jets ranges from  $50$ to $100$ depending on the number of tracks matching the jet candidate.

The missing transverse momentum is a two-dimensional vector defined as the  negative vector sum  of the  transverse momenta  of reconstructed electrons, muons, tau leptons and  jets, and also of calorimeter energy deposits not  associated with reconstructed objects.\footnote{Energy deposits in the calorimeters are expressed as four-vectors $(E,\mathbf{p})$, where the direction is determined from the position of the calorimeter cluster and the position of the primary vertex.}  The  magnitude  of the  missing transverse momentum vector is referred to as the missing transverse energy (\met).

\section{Event selection}
\label{sec:eventSel}

Events are required to be identified by the trigger system as containing at least one electron or one muon.  In order to control the data-taking bandwidth, the trigger system imposed a minimum transverse energy threshold on electrons of 20\,\GeV~or 22\,\GeV~(depending on the data-taking period), and a minimum  \pt~threshold on muons  of 18\,\GeV.  For the highest luminosities towards the end of the data-taking  period, the muon trigger is required  to be accompanied by  a jet that passes the first-level trigger \pt~threshold of 10 \GeV.  All data events are  required to be recorded during stable LHC running conditions  and with all relevant  sub-detectors functioning normally.  Events are cleaned for instrumental effects, such as sporadic noise bursts~\cite{EventCleaning}.

Events  are required to  have exactly one reconstructed  electron (muon) with \pt~$>$~25~(20)\,\GeV. This suppresses background processes such as \Zll~and \ttbar~which  have a higher average lepton multiplicity.  Exactly  one  identified   hadronic tau decay candidate with  \pt~$>$~30\,\GeV~and opposite-sign charge to the lepton  is required.  The \met~is  required to  be  larger  than 20\,\GeV~in order to further reject multi-jet and \Zll~processes.  In addition, at least two  reconstructed jets  are  required, with  the  leading jet  having \pt~$>$~50\,\GeV~and  the sub-leading  jet having  \pt~$>$~25\,\GeV.   

The signal-to-background ratio is improved  by requiring that either the leading or sub-leading jet passes the $b$-tagging requirements. After requiring that events must have one of these two jets passing the $b$-tagging requirements, the dominant background process is \ttbar.  Since both the signal and \ttbar~processes contain two $b$-jets in the final state, no further improvement in sensitivity is obtained by requiring that a second jet in the event also pass the $b$-tagging requirements.

The visible mass (\twomass{\tauhadvis}{-\mathrm{jet}}) of the tau candidate and the closest jet in $\eta  - \phi$  space (minimum  $\Delta R$) is required to be larger than 90\,\GeV.  Only jets with \pt~$>$~40\,\GeV~ are considered.   This cut is  chosen to reject  semi-leptonic \ttbar~events where the tau candidate is faked  by jets from $W \rightarrow  \qqbar$ decays.   

In events containing leptoquark decays, large \met~arises from neutrinos accompanying the tau decays.  Taus originating from leptoquarks typically have high momentum, thus the decay  products are predominantly collinear and the \met~direction is correlated with the direction of the visible tau decay products. Two variables are defined in order to improve the separation of signal and background: the absolute difference in $\phi$ between the charged lepton and \met~($|\Delta  \phi (\met, \ell) |)$, and the absolute difference in $\phi$ between the tau candidate and \met~( $|\Delta  \phi (\met, \tauhadvis) |$).  The relationship between these two variables for simulated signal ($m_{LQ}$ = 500\,\GeV) and the dominant top background, after applying  all the requirements described above is shown in figure~\ref{fig:METAngle}.  Events must satisfy the following requirement:
\begin{equation}
\label{eq:METAngleFunc}
|\Delta \phi (\met, \ell)| \le -\frac{1.5}{\pi} |\Delta \phi (\met, \tauhadvis)| + 2~\mathrm{(radians)},
\end{equation}
where $\ell=e,\mu$.  This \met~angular requirement selects events below the solid line in figure~\ref{fig:METAngle}. A leptonically decaying tau is accompanied by two neutrinos, whereas a hadronically decaying tau is accompanied by only one.  In events with two true taus (as in the signal process), the \met~is therefore typically aligned with the leptonic tau decay and these events are preferentially selected by the \met~angular requirement.  The signal efficiency is approximately 85\%, independent of leptoquark mass.  For \ttbar~events containing a real hadronic tau (produced from the \Wboson~decay), the additional neutrinos from the tau decay cause the \met~to be preferentially aligned with the hadronic tau decay and these are rejected.  In the subset of \ttbar~events where the tau is faked by a jet from $W$-decay, events are evenly distributed across the $|\Delta \phi (\met, \ell) |$-$|\Delta \phi (\met, \tauhadvis) |$~plane and a large proportion of these are also rejected.  The overall efficiency for inclusive~\ttbar~events is 31\%.

\begin{figure}
\centering
  \subfigure[Signal, $m_{LQ}$ = 500\,\GeV]{\label{fig:METAngleLQ} 
    \includegraphics[width=0.48\textwidth]{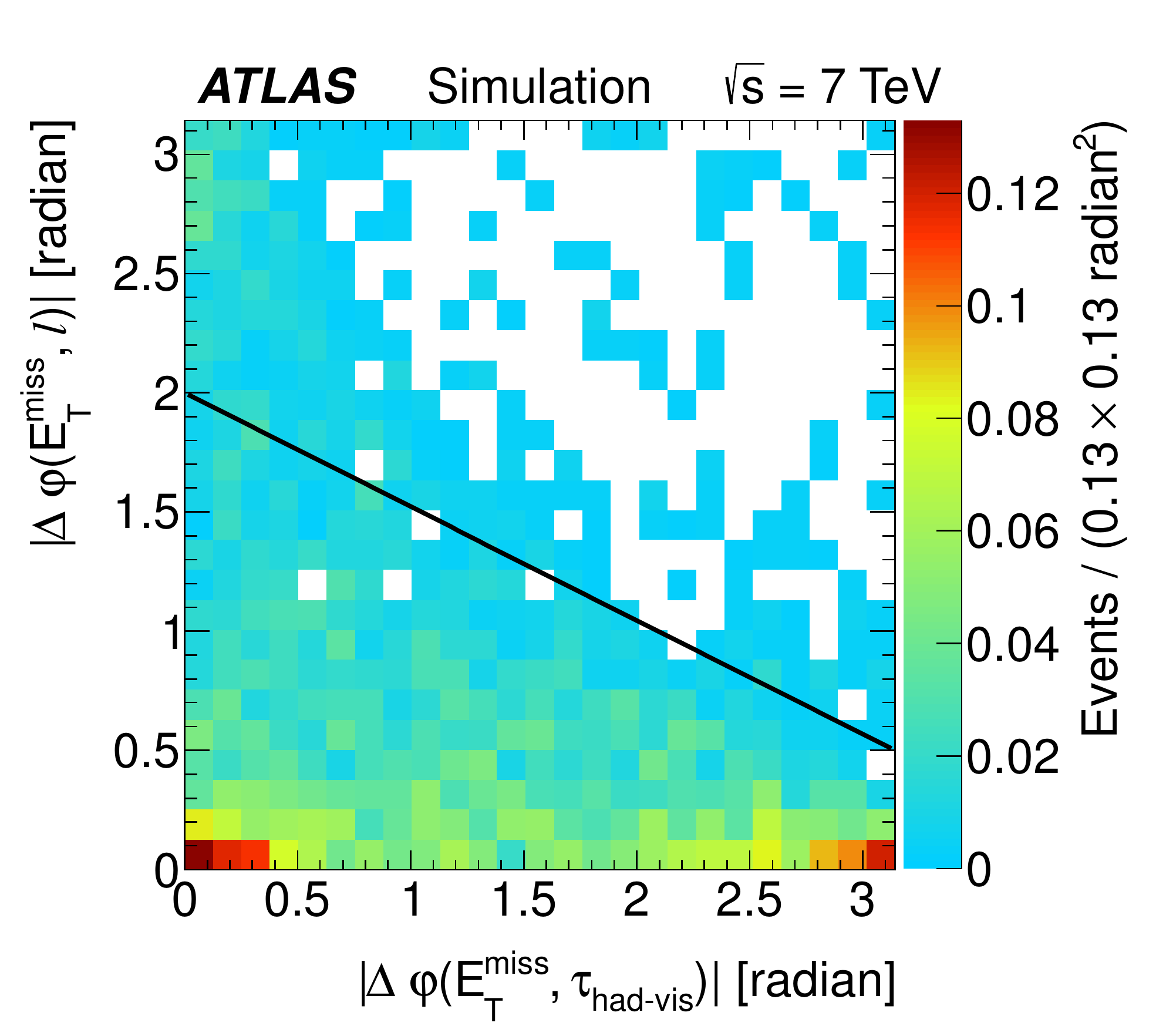} }
  \subfigure[Top-quark background]{\label{fig:METAngleTop}
    \includegraphics[width=0.48\textwidth]{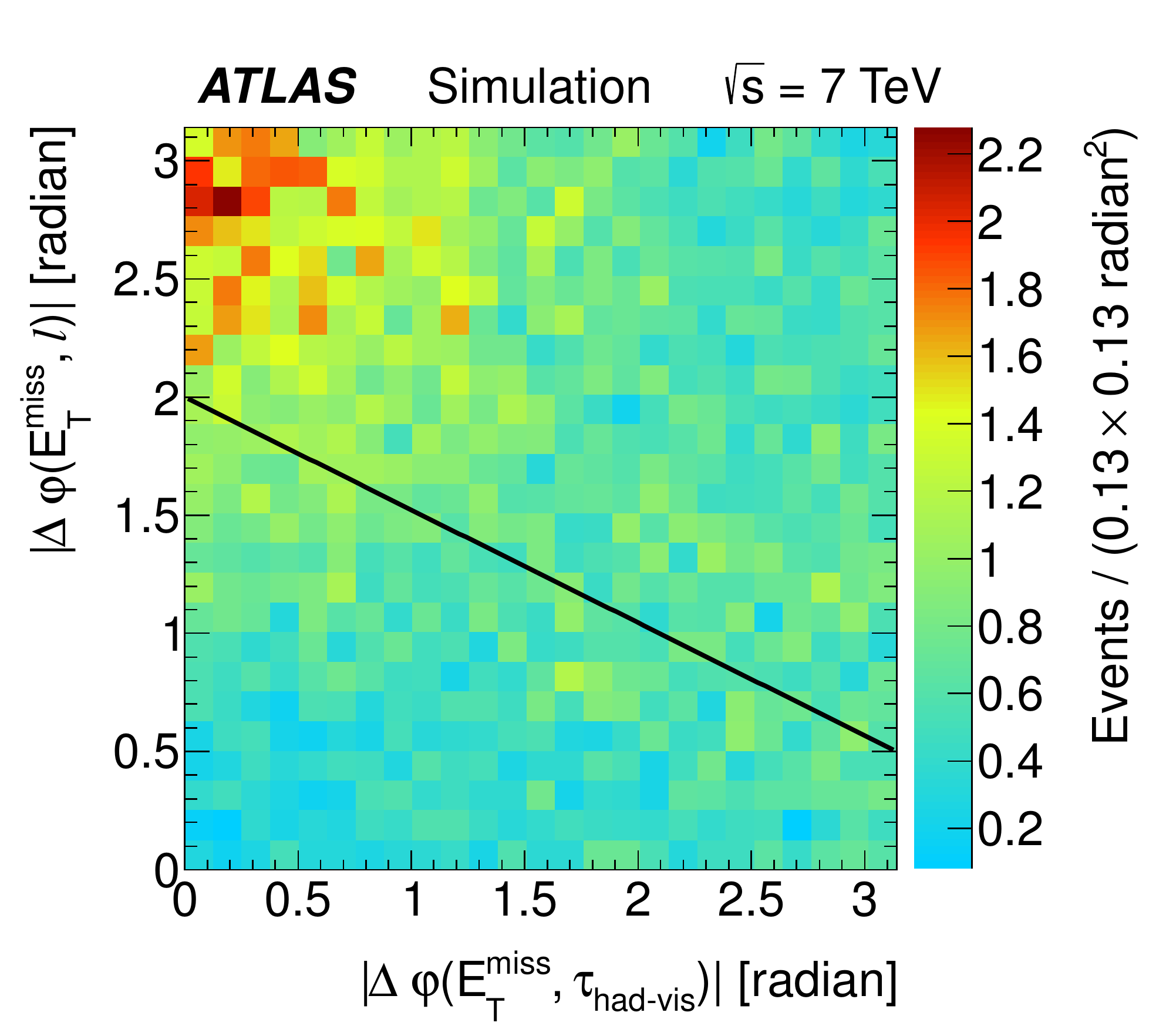}}
  \caption[]{ The  absolute value of the angle $\Delta \phi$ between the reconstructed  charged light  lepton and \met~as a function of $|\Delta  \phi|$ between \tauhadvis~and \met~for simulated \subref{fig:METAngleLQ} signal  ($m_{\LQ}$ = 500\,\GeV) and \subref{fig:METAngleTop} top-quark background, after applying all selection cuts (see text) and normalising to the integrated luminosity of the data.  The function corresponding to the solid line is defined in eq. \ref{eq:METAngleFunc}.}
  \label{fig:METAngle}
\end{figure}

\section{Background estimation}
\label{sec:BG}
Backgrounds considered in  this analysis are the production of \Wboson+jets and \ZGamma+jets (collectively referred to as $V$+jets), \ttbar, single top quarks, di-boson and multi-jets.  Normalisation factors are applied to the MC predictions for $V$+jets and top backgrounds in background-enriched control regions, to predict as accurately as possible the background in the signal region, as described in more detail below.  These control regions are constructed to be mutually exclusive of the signal region and the assumption is made that normalisation factors in the signal region are the same as in the background-enriched control regions.  The contribution from multi-jets is estimated using fully data-driven techniques.  The contribution from di-boson processes is taken directly from MC.  The shapes of the distributions are taken from simulation in the signal region. 

Different approaches are used to estimate the backgrounds in the electron and muon channels.  Normalisation factors for the electron channel are calculated after applying the electron, tau, \met, charge-product cuts, and jet multiplicity and \pt~requirements described in section \ref{sec:eventSel} above.  This approach minimises bias with respect to the signal region, but leads to limited statistics (for MC and data) in the control regions.  Normalisation factors for the muon channel are calculated after applying only the muon, tau and charge-product requirements described in section \ref{sec:eventSel}. 

\subsection{Electron channel}
\label{sec:BGElec}

The multi-jets background is estimated by defining a region in data with a tau candidate that fails the tau BDT identification used in the nominal selection but passes a looser identification working point, and has the same-sign charge as the electron.  In addition, the events are required not to contain any taus that  pass the  nominal selection criteria.  Contributions  from \Wboson,
\ZGamma+jets,  top-quark,  and  di-boson   background  processes estimated from  MC simulations  are  subtracted  to   get  the  shape  of  the  multi-jets distribution.   The normalisation is  determined by performing a two-component maximum likelihood fit to kinematic distributions of the sum  of  multi-jets  and  all other  sources  of background  to  data,  with  the  multi-jets  fraction  being  the  fit parameter.  The variable used for fitting is chosen  to provide good discrimination between multi-jets and other sources of background.  The method  is used  to calculate the multi-jets contribution in  the signal region, where the transverse mass between the charged light lepton and the \met, defined to be:
\begin{equation}
m_{\mathrm{T}} = \sqrt{2\pt^{\mathrm{\ell}} \met(1-\cos (\Delta \phi))},
\label{eq:mT}
\end{equation}
\noindent is used as the fit variable.  The multi-jets contribution is found to be 12$^{+8}_{-16}$\% of the total data yield in the signal region.  The same method is also used in background control regions, fitting to the \met~distribution in the \Wboson~and \ZGammatau~control regions, and the electron \ET~in the top control region.  The validity of the method used to estimate the multi-jets background contribution is cross-checked by using events with same-sign charge electron--tau pairs as the control region and found to be compatible within statistical errors.  Dependence on the choice of variable used is evaluated by fitting to other kinematic variables, and also found to be within statistical errors of the nominal value.

Separate  control regions  are defined  for the \ZGammaee, \ZGammatau, \Wboson, and \ttbar~and single top-quark processes.  They are defined by applying the electron, tau, charge-product, \met, and jet multiplicity and \pt~requirements (as described in section~\ref{sec:eventSel}, collectively referred to as the `baseline' requirements), in addition to the  cuts  shown  in  table~\ref{tab:CRElec}.  

The control region for \ZGammaee~events is constructed by requiring an additional electron with \pt~$>$~20\,\GeV~and opposite-sign charge to the first one.  The second electron is required to pass the same identification requirements as the first one.

The \ZGammatau~control region is defined by additionally requiring that the visible mass of the electron--tau pair is in the range 40~$<$~\twomass{e}{\tauhadvis}~$<$~80\,\GeV~and that the transverse mass between the electron and the \met~is less than 60\,\GeV.  A $b$-jet veto is also applied, using a looser working point (with a selection efficiency of 75\%) compared to the working point used for signal selection.  In this way contamination from top backgrounds is reduced.

The \Wboson+jets control region is constructed by applying in addition a $b$-jet veto (as described above), demanding that 60~$< m_{\mathrm{T}} <$ 120\,\GeV, and requiring that the event fail the \met~angular requirements cut (eq. \ref{eq:METAngleFunc}).

The control region for top backgrounds is defined by additionally requiring that events pass the $b$-tagging requirements, have $\twomass{\tauhadvis}{jet} > $ 90\,\GeV, fail the \met~angular requirements and have \ST $ < $ 350\,\GeV, where  \ST~is defined as the scalar sum of the \pt~of the charged light lepton, the tau, the two highest-\pt~jets and the \met~in the event,
\begin{equation}
\ST = \pt^{\mathrm{e/\mu}}+\pt^{\mathrm{\tauhadvis}}+\pt^{\mathrm{jet1}}+\pt^{\mathrm{jet2}}+\met.
\end{equation}

\begin{table}
\begin{center}
\begin{tabular}{|c|c|} \hline
\ZGammaee+jets                             & Require one extra electron                    \\ \hline
\multirow{3}{*}{\ZGammatau+jets}           & $b$-jet veto                                  \\
                                           & 40~$<$ \twomass{e}{\tauhadvis} $<$ 80\,\GeV   \\ 
                                           & $m_{\mathrm{T}}$ $<$ 60\,\GeV                  \\ \hline
\multirow{2}{*}{$W$+jets}                       & $b$-jet veto                                  \\ 
                                           & Fail \met~angular  cut                        \\
                                           & 60~$< m_{\mathrm{T}} <$ 120\,\GeV               \\ \hline
\multirow{4}{*}{Top-quark}                 & $b$-jet requirement                           \\
                                           & $\twomass{\tauhadvis}{jet} > $ 90\,\GeV       \\ 
                                           & Fail \met~angular cut                         \\                                         
                                           & $\ST < $ 350\,\GeV                            \\ \hline
\end{tabular}
\caption{Control region definitions for the electron channel.  The electron, tau, charge-product, \met, and jet multiplicity and \pt~requirements are applied as described in section~\ref{sec:eventSel}.}
\label{tab:CRElec}
\end{center}
\end{table}

Normalisation factors for $V$+jets and top backgrounds in the electron channel are calculated according to:
\begin{equation}
NF_{\mathrm{BG}} = \frac{N^{\mathrm{Data}} - N^{\mathrm{MC}}_{\mathrm{Other BG}}}{N^{\mathrm{MC}}_{\mathrm{BG}}},
\label{eq:SF}
\end{equation}
where $N^{\mathrm{Data}}$ is  the number of data events  in the control region, $N^{\mathrm{MC}}_{\mathrm{Other~BG}}$ is the expected number of events from other background processes, and $N^{\mathrm{MC}}_{\mathrm{BG}}$ is the contribution in the control region from the background process of interest.  

To account  for contamination from  other background processes in the control regions, normalisation  factors are found for each region in turn.  At each stage the multi-jets contribution is re-estimated and all previously found  normalisation factors are applied when estimating  the contribution from other background processes. 

The final background normalisation factors obtained are presented in table~\ref{tab:factors}~and discussed together with the muon channel in section \ref{sec:BGSummary}.

\subsection{Muon channel}
\label{sec:BGMuon}
The  multi-jets contribution to the  control and  signal  regions are
estimated in the muon channel from data using the ABCD method. Events are sorted
into  four  regions  using  two  observables assumed to be independent  --  the  muon
isolation and  the sign  of the charge  product of the  muon--tau pair.
The regions are therefore  defined as: isolated muon and opposite-sign
muon--tau pair ($A$), isolated  muon and same-sign muon--tau pair ($B$),
as well as  two regions with a non-isolated  muon and opposite-sign or
same-sign  charge muon--tau  pair  ($C$  and  $D$  respectively).  Non-isolated muons are defined as those which fail at least one of the isolation requirements described in section \ref{sec:eventSel}.
Contributions from other physics processes in regions $B$, $C$,
and $D$~are subtracted using the MC simulation.  The shape of kinematic distributions for  the multi-jets background is taken  from region $B$,
while  the expected number  of events  in the  signal region  ($A$) is
determined by  taking the  product of the  number of events  in region
$B$~with  the ratio  of the  number of  events in  regions $C$~and
$D$~({\it i.e.} $N_A = N_B \times \frac{N_C}{N_D}$).  The multi-jets contribution is estimated to be 15$\pm$4\% of the total data yield in the signal region.  The validity of the method is checked by varying both isolation cuts up and down from the nominal value of 0.2 by 0.05.  Deviations in the ratio $\frac{N_C}{N_A}$~are included as an additional systematic uncertainty. 

Control regions for $V$+jets and top-quark background processes are defined  by applying the muon and tau requirements (including the charge-product requirement) as described in section~\ref{sec:eventSel}.   Additional selection criteria used  for each  control region  are listed  in table~\ref{tab:CRMuon}.   Normalisation factors are calculated for each process by performing a maximum likelihood fit.  The variable used for fitting is chosen in each case to provide good discrimination between the background process of interest, and other contributing physics processes in that control region.  The contribution from multi-jets in each control region is estimated using the method described above, and this and contributions from other background processes are taken into account when performing the fits.

The control  region for \ZGammamm~events  is defined by  requiring two oppositely charged  muons and  one hadronic tau decay.   The second  muon is required to  pass the same  requirements as the first.  The normalisation factor for \ZGammamm~events in the signal region is then determined by fitting to the di-muon invariant mass distribution in the range 60~$<$~\mass{\mu}~$<$~120~\GeV.

The  normalisation  of \ZGammatau~events is obtained by selecting events with one muon, one hadronic tau decay and \met~$>$~20\,\GeV.  Additionally, events are required to fail the $b$-jet requirement.   The fit is performed using the visible mass of the muon--tau pair in the range 45~$<$~\twomass{\mu}{\tauhadvis}~$<$~80\,\GeV.

The control region for \Wboson+jets events is defined by selecting events with one muon, one hadronic tau decay and \met~$>$~20\,\GeV, and which fail the $b$-jet requirement.  The normalisation is determined by fitting to the transverse mass of the charged light lepton and the \met~in the range 70~$<$~$m_{\mathrm{T}}$~$<$~100\,\GeV.  

The control  region for \ttbar~and  single-top processes  is defined by applying  all selection  criteria,  with the  exception  of the  \met~angular requirement which is reversed.  In addition, the \ST~of the event is required to be less than 350\,\GeV.  The normalisation factor is obtained by fitting to the \ST~distribution up to  350\,\GeV.  

\begin{table}
\begin{center}
\begin{tabular}{|c|c|} \hline
\multirow{2}{*}{\ZGammamm+jets}            & Require one extra muon                                 \\ 
                                           & 60~$<$~\mass{\mu}~$<$~120\,\GeV                        \\ \hline
\multirow{3}{*}{\ZGammatau+jets}           & \met~$>$~20\,\GeV                                      \\
                                           & $b$-jet veto                                           \\
                                           & 45~$<$~\twomass{\mu}{\tauhadvis}~$<$ 80\,\GeV          \\ \hline
\multirow{2}{*}{$W$+jets}                  & \met~$>$~20\,\GeV                                      \\
                                           & $b$-jet veto                                           \\
                                           & 70~$< m_{\mathrm{T}} <$ 100\,\GeV                        \\ \hline
\multirow{4}{*}{Top-quark}                 & \met~$>$~20\,\GeV                                       \\
                                           & $\pt^{\mathrm{jet1}}$~$>$~50\,\GeV                        \\
                                           & $b$-jet requirement                                     \\ 
                                           & $\twomass{\tauhadvis}{-\mathrm{jet}} > $ 90\,\GeV       \\ 
                                           & Fail \met~angular cut                                   \\
                                           & $\ST < $ 350\,\GeV                                      \\ \hline
\end{tabular}
\caption{Control region definitions for the muon channel.  The muon, tau and charge-product requirements are also applied as described in section~\ref{sec:eventSel}.}
\label{tab:CRMuon}
\end{center}
\end{table}

\subsection{Background summary}
\label{sec:BGSummary}

The background normalisation  factors in the signal region determined from data for both channels are presented in table~\ref{tab:factors}.  

\begin{table}
\begin{center}
\begin{tabular}{|l|c|c|} \hline
Background                   & Electron channel                         & Muon channel        \\\hline
\ZGammallMargar+jets        & 0.54 $\pm$ 0.09                          & 0.52 $\pm$ 0.02     \\
\ZGammatau+jets             & 0.99 $\pm$ 0.08                          & 1.00 $\pm$ 0.02     \\
\Wboson+jets                & 0.63 $\pm$ 0.07                          & 0.50 $\pm$ 0.01     \\
\ttbar~and single-top       & 0.92 $\pm$ 0.08                          & 0.93 $\pm$ 0.09     \\\hline
\end{tabular}
\caption{Summary of background normalisation factors obtained using the control regions specified in tables \ref{tab:CRElec}~and \ref{tab:CRMuon}.  The errors include both the statistical and systematic (discussed in section \ref{sec:syst_bkg}) uncertainties.}
\label{tab:factors}
\end{center}
\end{table}

Uncertainties for normalisation factors are larger in the electron channel than the muon channel due to the tighter requirements placed on the control region definitions, namely the additional requirements on \met~and jets which are not applied for the muon channel (unless explicitly stated).

As a cross-check, the electron channel method (detailed in section \ref{sec:BGElec}) is applied to the muon channel and the signal region normalisation factors determined in this way are found to be 
$NF_{Z/\gamma \rightarrow \mu\mu}$ = 0.59 $\pm$ 0.09, 
$NF_{Z/\gamma \rightarrow \tau\tau}$ = 1.03 $\pm$ 0.08, 
$NF_W$ = 0.50 $\pm$ 0.08 and  
$NF_{\mathrm{top}}$ = 0.93 $\pm$ 0.09.  
These figures agree within uncertainties with the factors determined using the method described in section \ref{sec:BGMuon}~and shown in table \ref{tab:factors}.

The largest background contribution comes from \ttbar~events, with approximately 55\% of these coming from events containing a real hadronic tau decay (from the \Wboson~decay) after all selection cuts are applied.  Approximately 40\% come from events where the \Wboson~boson decays hadronically, and the tau candidate is faked by one of the jets.  In the remaining $\sim 5\%$ of events, the reconstructed hadronic tau decay is faked in equal proportions by electrons or $b$-jets.  Normalisation factors for background processes in which the hadronic tau decay is faked by a jet are observed to be significantly smaller than unity.  This is a known effect, caused by jets being narrower in simulation than in data and therefore being more likely to fake a hadronic tau decay~\cite{tauMisID}.

In both  methods used for  background estimation, the  control regions for $V$+jets background processes either make no requirements on $b$-tagging, or veto events containing one or more $b$-jets.  Simulation tests have validated the assumption that the normalisation factors obtained in regions that require $b$-jets are the same as those in regions where $b$-jets are not explicitly required, or are vetoed.

\section{Systematic uncertainties}
\label{sec:syst}

In simulated samples all sources of uncertainty are varied individually within their errors and the impact on the results of the analysis is determined.  Background normalisation factors and multi-jets contributions are recalculated for each source of systematic uncertainty.  In this way the nominal simulation (comprising the current best estimates for physics object reconstruction corrections) and systematic variations are treated coherently, and the uncertainties are propagated through the analysis.  The \ST~distribution is used to test for the existence of leptoquarks, since this variable provides the best discrimination between signal and background (discussed further in section \ref{sec:results}). The shape of the \ST~distribution remains unchanged within the total shape uncertainty (see further discussion in section \ref{sec:results}) when applying all the uncertainties detailed below and systematic variations are therefore treated as nuisance parameters affecting the overall scale.  The relative changes in acceptance for signal and background are quoted for each systematic variation.

\subsection{Object-level uncertainties}

There  are  several  sources  of systematic uncertainty  associated with the reconstruction, identification, and energy scale of physics objects, which can potentially affect the estimated shapes of kinematic distributions and the normalisation of various processes. 

Uncertainties associated with the efficiency of single-lepton triggers are typically less than 1\%~\cite{ElectronTrigger,MuonTrigger}~and a $\pm$0.5\% ($\pm$3.3\%) variation in the signal acceptance is observed when varying the electron (muon) trigger efficiency by $\pm$1$\sigma$.  The effect on background processes is negligible. 

Varying  the electron  energy scale by $\pm$ 1$\sigma$
results in  a $\pm$0.8\% change  in background acceptance  compared to
the  nominal selection and has a negligible impact on the signal yields.   The
electron reconstruction and identification  efficiency   uncertainties  are
combined in quadrature and yield  an overall change of less than 1.5\%
for both signal and background.

Varying the muon momentum scale by 1$\sigma$ results in a 0.2\% change in signal yields
compared  to   the  nominal  selection.   The   impact  of muon resolution
uncertainties  on signal  and background  acceptance is  found  to be
negligible.   A   $\pm$1$\sigma$  variation  of   muon  reconstruction
efficiency results in a $\pm$0.3\% change in signal acceptance.

The uncertainty on the tau energy scale is typically around 3\%, depending on the \pt~and $\eta$ of the hadronically decaying tau candidate \cite{tauEnergyScale}.  Varying the energy scale by $\pm$1$\sigma$ changes the acceptance for background and low-mass signal ($m_{\LQ}$~=~200\,\GeV) by approximately 2\%, decreasing to 1.2\% for $m_{LQ}$~=~500\,\GeV.  The uncertainty on the tau identification efficiency is 4\% for taus with \pt~$<$ 100 GeV.  This increases linearly with \pt~up to a maximum of 8\% for three-prong taus with \pt~$>$ 350 \GeV.  Varying the tau identification efficiency by this uncertainty results in an overall acceptance change for signal of approximately 6\% (for a leptoquark of mass 500 \GeV).  The variation of background yields is found to be approximately 1\% -- significantly smaller than the change in the signal yield, because the effect is largely absorbed in the normalisation factor defined in the control region.

The jet energy resolution uncertainty is approximately 10\% and affects the event yields
by approximately 2\%~\cite{JES}. The jet energy scale uncertainty depends on \pt~and $\eta$, and varies between 2\%
and 5\%.  It is modelled by 14 separate nuisance parameters, each of which is varied by $\pm$1$\sigma$ independently from the others. 
The use of control regions does not significantly reduce the variations of the different background yields, and changes in acceptance of signal and background of up to $\pm$2\% are observed.  

The uncertainty  on the $b$-jet identification efficiency for the algorithm and operating point used in this analysis ranges from 5\% to 18\% depending on jet kinematics.  The $b$-tagging efficiency and probability that a light jet is identified as a b-jet are anti-correlated and thus varied accordingly.  A $\pm$1$\sigma$  variation results in a $\pm$9\% ($\pm$15\%) change in signal acceptance for leptoquarks with a mass of 200~(500)\,\GeV.  The use of control regions reduces the background yield variation to $\pm$3\% in both channels.  

All energy scale and resolution corrections for electron, muon, tau and jet candidates are propagated consistently to the \met~calculation.  Additional uncertainties related to the energy scale and pile-up dependence of calorimeter clusters not associated with any high-\pt~objects  (electrons, taus, jets) are also considered in the \met~calculation.  These sources are varied independently within their uncertainties and the impact on signal and background yields is found to be negligible.

The uncertainty on the integrated luminosity for data taken during 2011 is $\pm$3.7\% as determined in ref.~\cite{lumi}.

\subsection{Theoretical uncertainties}

QCD renormalisation and factorisation scales are varied by a factor of two to estimate the impact on the signal production cross-section.  The variation is found to be $\pm$12\%.  A re-weighting technique is used to assess the sensitivity of the signal acceptance to the choice of parton distribution functions and the resulting uncertainty is estimated  to be  $\pm$12\%.  Varying the multi-parton interactions within experimental bounds has a negligible effect on the signal process.

The effect of the choice of event generator for the top-quark background is estimated by using PowHeg 1.0~\cite{PowHeg, PowHeg1} (instead of MC@NLO 4.01) to model the hard process.  The parton shower and hadronisation models, and the underlying event model (respectively HERWIG 6.510 and JIMMY 4.31 in the nominal sample) are replaced with those from PYTHIA 6.425.  In addition, the CTEQ6L1 PDF set is used instead of CT10 which is used in the nominal \ttbar~sample.  The total background yield is found to differ by 1.5\% with respect to the nominal samples.  

The uncertainty on the signal and the top-quark background due to initial-state radiation (ISR) and final-state radiation (FSR) is evaluated using the AcerMC generator interfaced to the PYTHIA 6.425 shower model with the parameters controlling ISR and FSR varied in a range consistent with experimental data~\cite{det1A}. The event yields are found to agree with nominal values within statistical uncertainties.

MC@NLO events with top-quark masses of 170 GeV and 175 GeV are used to assess the top-quark mass dependence, which is added in quadrature to the uncertainty related to the choice of event generator and PDF set.  The resulting uncertainty ($2.8\%$) is treated as a nuisance parameter affecting the background yield and is assumed to be fully correlated between the electron and muon channels. 
 
Other background processes taken from simulation (\Wboson, \ZGamma,  di-boson) account for less than 20\% of the total background.  The \Wboson~and \ZGamma~samples are simulated with the matching parameter (described in ref.~\cite{MLM}) set  to 20\,\GeV.  Event yields are found to agree with the nominal values within statistical uncertainties when this parameter is changed to 30~\GeV.

The \Wboson~and \Zboson~control regions are defined with either the application of a $b$-jet veto, or with no $b$-tagging requirements.  The  uncertainties on the production cross-sections of \Wboson~or \Zboson~bosons in association with one or two $b$-jets are estimated using MCFM~\cite{MCFMWbb, MCFMZbb}.  The QCD renormalisation and factorisation scales are varied independently by a factor of two, and different PDF sets are considered.  The total uncertainty is found to be 30\% and the uncertainties on the normalisation factors for \Wboson~and \Zboson~background processes are increased by this amount ({\it i.e.} to 2\%).

\subsection{Background uncertainties}
\label{sec:syst_bkg}

For each channel, the systematic  uncertainties on the normalisation factors in table \ref{tab:factors} are evaluated by calculating the normalisation factor for a given background when normalisation factors for all other sources of background are varied up or down by their statistical error. The systematic uncertainty on the multi-jets background is evaluated by varying the normalisation factors of other backgrounds by $\pm 1 \sigma$.  The methods used to estimate the contributions from multi-jets processes are validated by modifying the control regions used, and in the case of the electron channel the variable used for fitting (see sections \ref{sec:BGElec}~and \ref{sec:BGMuon}).  Deviations from the nominal value are included as additional sources of systematic uncertainty.

Conservatively, all background normalisation  factors are assumed to be fully correlated and the impact on the total background yield is $^{+16}_{-19}\%$ for the electron channel and $\pm9\%$ for the muon channel. The background estimation method used in the muon channel allows a more accurate determination of the normalisation factors compared to the event-counting method employed in the electron channel, and the normalisation factor uncertainty for the muon channel is correspondingly smaller than in the electron channel.  For both channels, the limited number of data events in the top-quark control region is the main source of uncertainty on the top-quark normalisation factor, which in turn has the largest impact on the total yield uncertainty.  Due to the tighter requirements on control regions for the electron channel background estimation, this method also suffers from a limited number of events in data and simulation when estimating normalisation factors for $V$+jets background processes. 

\subsection{Summary of systematic uncertainties}
  
The shape of the \ST~distribution remains unchanged within statistical uncertainties when applying all the uncertainties mentioned above.  
The uncertainties for the electron and muon channels are summarised in tables~\ref{tab:syst_el}~and~\ref{tab:syst_mu}~respectively.  Uncertainties related to the background normalisation factors have the largest impact on the total background yield, while for the signal yield the largest sources of systematic uncertainty are due to theoretical uncertainties (comprising uncertainties related to PDFs, multi-parton interactions, and initial- and final-state radiation) and from $b$-jet identification. 

\begin{table}
\centering
\begin{tabular}{|c|c|c|c|}
\hline
                                                       &    Background     &      $\LQ$($m$=500\,\GeV)   \\ \hline
                        Luminosity                     &   $      -    $   &      $      3.7 $   \\ \hline
                            Theory                     &   $      2.8  $   &      $       17 $   \\ \hline
             Normalisation factors                     &   $   +16/-19 $   &      $       -  $   \\ \hline
                Trigger efficiency                     &   $   < 0.2   $   &      $      0.5 $   \\ \hline
             Electron energy scale                     &   $     0.8   $   &      $      0.4 $   \\ \hline
Electron reconstruction and identification efficiency  &   $     1.5   $   &      $      1.5 $   \\ \hline
                  Tau energy scale                     &   $     2.6   $   &      $      1.2 $   \\ \hline
                 Tau ID efficiency                     &   $     1.2   $   &      $      5.7 $   \\ \hline
   Jet energy scale (nuisance parameter dependent)     &   $ 0.1-2.4   $   &      $    < 0.2 $   \\ \hline
             Jet energy resolution                     &   $      2.5  $   &      $      0.3 $   \\ \hline
            $b$-tagging efficiency                     &   $      2.7  $   &      $       15 $     \\ \hline
\end{tabular}
\caption{The sources of systematic uncertainty in the electron channel and the relative change (in \%) in the background and signal yields.  The theory term includes  uncertainties related to initial and final state radiation, PDFs, and multi-parton interactions.}
\label{tab:syst_el}
\end{table}

\begin{table}
\centering
\begin{tabular}{|c|c|c|c|}
\hline
                                                   &    Background     &      $\LQ$($m$=500\,\GeV)   \\ \hline
                     Luminosity                    &   $      -    $   &      $      3.7 $   \\ \hline
                         Theory                    &   $      2.8  $   &      $       17 $   \\ \hline
          Normalisation factors                    &   $       9   $   &      $       -  $   \\ \hline
             Trigger efficiency                    &   $   < 0.2   $   &      $      3.3 $   \\ \hline
            Muon momentum scale                    &   $     0.1   $   &      $      0.2 $   \\ \hline
   Muon reconstruction efficiency                  &   $    < 0.1  $   &      $      0.3 $   \\ \hline
               Tau energy scale                    &   $     0.6   $   &      $      1.2 $   \\ \hline
              Tau ID efficiency                    &   $     0.8   $   &      $      5.7 $   \\ \hline
Jet energy scale (nuisance parameter dependent)    &   $   0.1-2.0 $   &      $    < 0.2 $   \\ \hline
          Jet energy resolution                    &   $      1.5  $   &      $      0.5 $   \\ \hline
         $b$-tagging efficiency                    &   $      2.7  $   &      $       15 $   \\ \hline
\end{tabular}
\caption{The sources  of systematic uncertainty in  the muon channel and the relative change (in \%) in the background and signal yields.  The theory term includes uncertainties related to initial- and final-state radiation, PDFs, and multi-parton interactions. }
\label{tab:syst_mu}
\end{table}

\section{Results}
\label{sec:results}

The observed  yields in  data, as well as expected yields for the background processes and the signal for several leptoquark masses, after all selection cuts are applied, are shown in table~\ref{tab:yields}.  The \ST~distribution is used to test for the existence of leptoquarks.  Distributions for both channels are shown in figure~\ref{fig:ST}.

\begin{figure}
 \centering
\subfigure[Electron Channel]{\label{fig:st_final_elec}\includegraphics[width=0.49\textwidth]{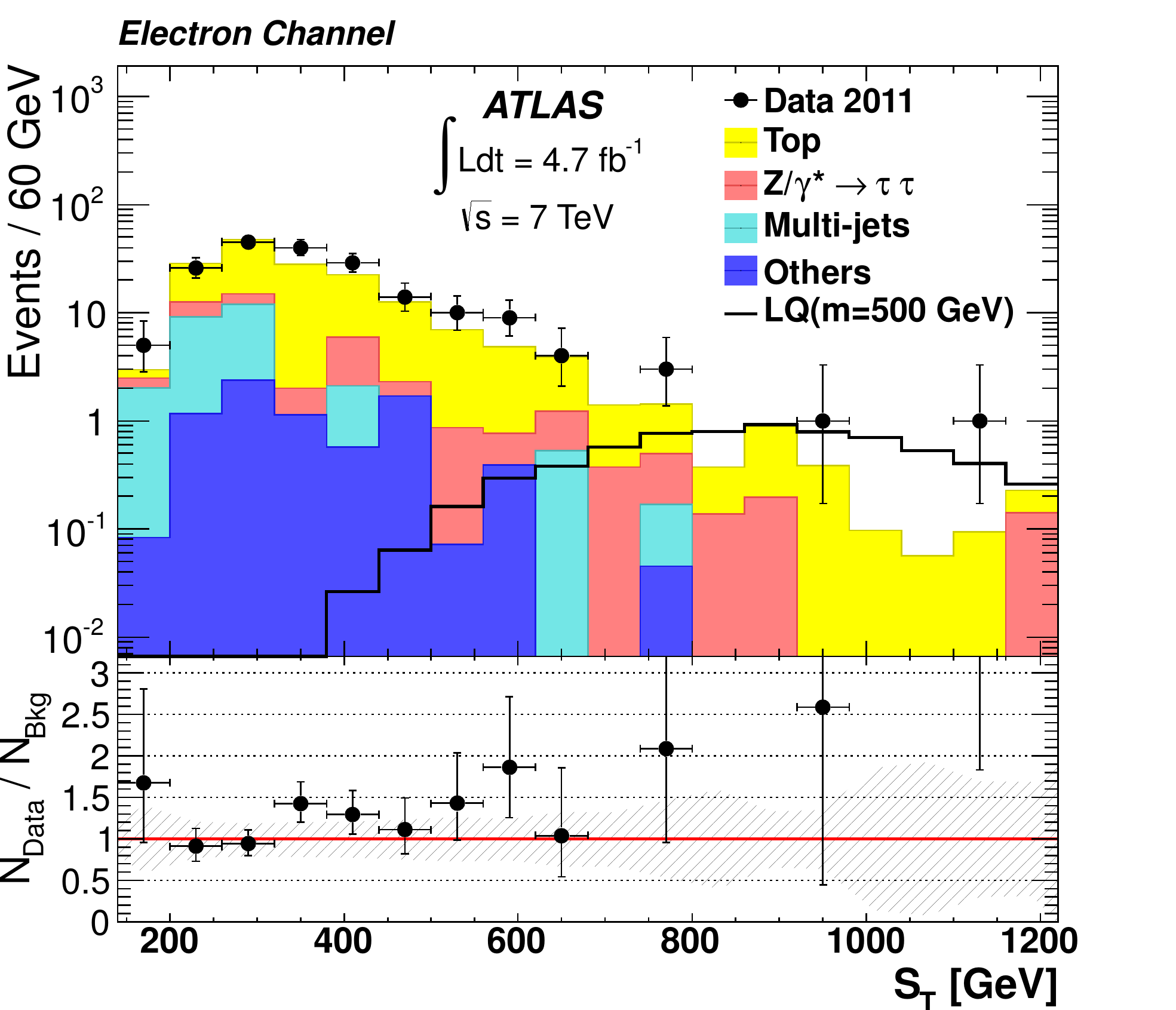}}
\subfigure[Muon Channel]{\label{fig:st_final_muon}\includegraphics[width=0.49\textwidth]{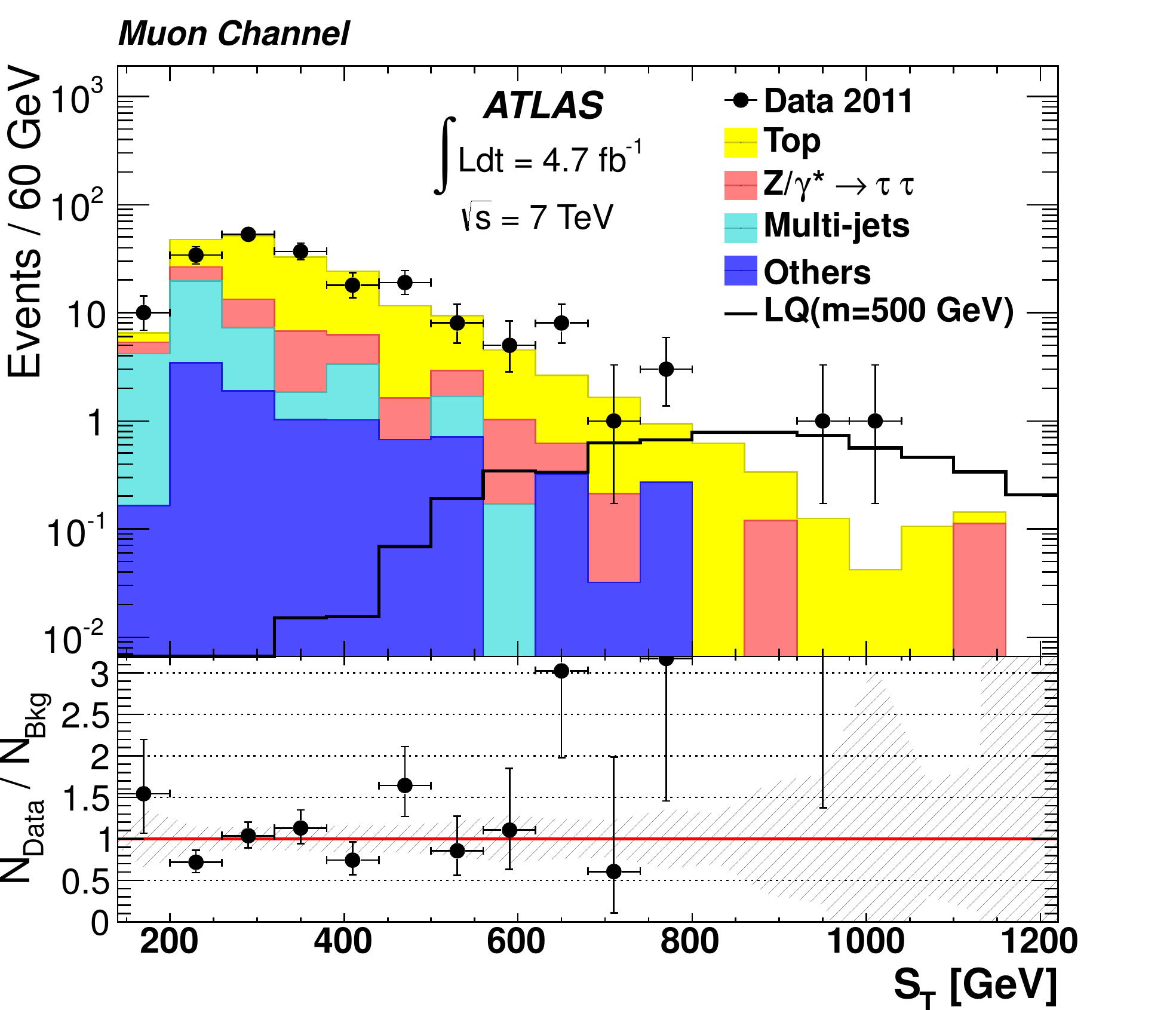}}
\caption[]{Data  and  MC comparisons  of  the  \ST~variable  after applying  all cuts in the \subref{fig:st_final_elec} electron and \subref{fig:st_final_muon} muon channels.  The ratio $N_{\mathrm{Data}}/N_{\mathrm{Background}}$ is also shown, where the red line at unity and hashed band represent the Standard Model expectation and associated statistical and systematic uncertainties.  No events with \ST~$>$~1.2\,\TeV~were observed in data.}
\label{fig:ST}
\end{figure}

\begin{table}
\begin{center}
\begin{tabular}{|l|c|c|} \hline
                        & Electron channel          & Muon channel       \\ \hline   
\ZGammallMargar+jets    & 2.1 	$\pm$ 0.9           & 1.1   $\pm$ 0.5    \\               
\ZGammatau+jets         & 21.4	$\pm$ 2.2	    & 26.6  $\pm$ 2.2    \\ 
\Wboson+jets 	        & 5.2	$\pm$ 1.2	    & 7.1   $\pm$ 1.5    \\                             
\ttbar~and single-top	& 119	$\pm$ 11	    & 130   $\pm$ 11     \\               
Di-boson	        & 1.1	$\pm$ 0.2	    & 1.3   $\pm$ 0.2    \\               
Multi-jets              & 20	$^{+13}_{-16}$       & 29    $\pm$ 8       \\ \hline   
Total background        & 169   $^{+27}_{-32}$       & 195   $\pm$ 18      \\ \hline         
Data	                & 187 	                    & 198                \\ \hline           
$m_{\LQ}$ = 200\,\GeV	& 711	$\pm$ 22	    & 839   $\pm$ 25     \\               
$m_{\LQ}$ = 300\,\GeV	& 131	$\pm$ 3	            & 136   $\pm$ 3      \\               
$m_{\LQ}$ = 400\,\GeV	& 28.7	$\pm$ 0.6	    & 28.6  $\pm$ 0.6    \\               
$m_{\LQ}$ = 500\,\GeV	& 7.53	$\pm$ 0.15	    & 6.84  $\pm$  0.15  \\               
$m_{\LQ}$ = 600\,\GeV	& 2.1	$\pm$ 0.04          & 1.87  $\pm$  0.04  \\  \hline
 
\end{tabular}
\caption{Yields  for data, background  and several  leptoquark masses in both channels after all cuts are applied. The errors include statistical uncertainties and systematic uncertainties on the background normalisation.}
\label{tab:yields}
\end{center}
\end{table}		   

At very high \ST, the statistical uncertainties on the various background processes become very poor due to the limited number of MC and (in the case of the multi-jets) data  events in the signal region.  The sum of the background processes is fitted in the region 350~\GeV~$<$~\ST~$<$~2000\,\GeV~to an exponential function using a maximum likelihood fit.  In this way the distribution is smoothed and a background expectation is provided throughout this \ST~region.  The fit  parameters are varied  within their uncertainties to obtain a shape uncertainty.  The shape of the \ST~distribution is checked for all systematic variations and the variation is found to be significantly smaller than the fit uncertainty in almost all cases.  The only exception is for the variation in choice of generator used to model the \ttbar~background process, where the central value lies outside the nominal range (although covers the nominal value within its own statistical uncertainty).  Conservatively, the difference between the nominal shape and the alternative shape is taken as a systematic uncertainty and added to the shape uncertainty determined from the nominal fit. The total shape uncertainty is treated as an additional nuisance parameter. Comparisons of the fitted distributions to data are shown in figure~\ref{fig:st_fitted}.  Below \ST~=~350\,\GeV~the background shape is taken from the histogram.

\begin{figure}
  \centering
  \subfigure[Electron Channel]{\label{fig:st_fitted_elec}\includegraphics[width=0.49\textwidth]{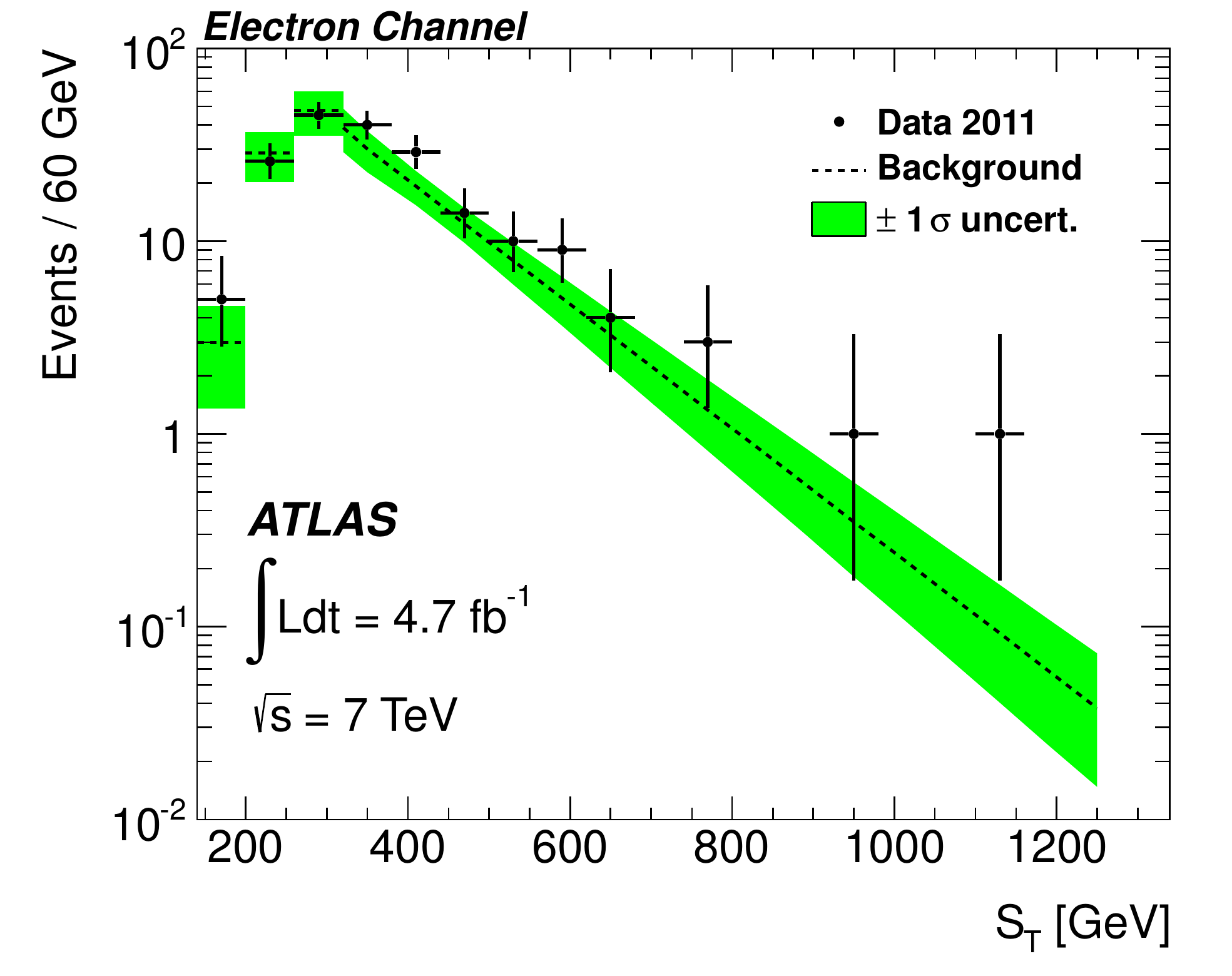}}
  \subfigure[Muon Channel]{\label{fig:st_fitted_muon}\includegraphics[width=0.49\textwidth]{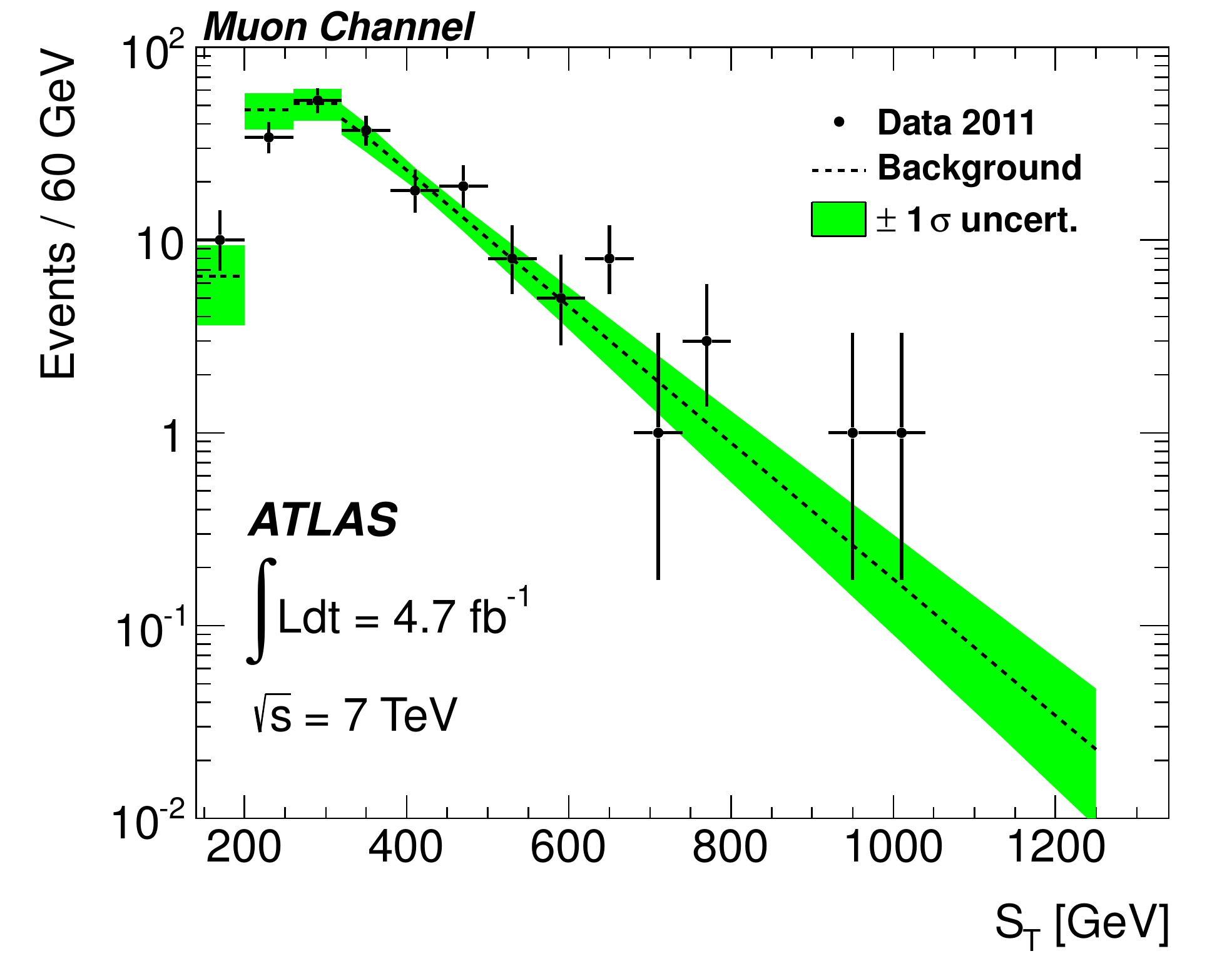}}
  \caption[]{Comparison  of the fitted  \ST~background  shape to  data in the \subref{fig:st_final_elec} electron and \subref{fig:st_final_muon} muon channels.  The $\pm 1 \sigma$~band represents the uncertainty on the shape of the \ST~distribution, obtained by varying the fit parameters within their uncertainties and comparing with the shape of the \ST~distribution obtained for each systematic variation.  No events with \ST~$>$~1.2\,\TeV~were observed in data.}
  \label{fig:st_fitted}
\end{figure}

\begin{figure}
\centering
\subfigure[Electron Channel]{\label{fig:limit_final_elec}\includegraphics[width=0.49\textwidth]{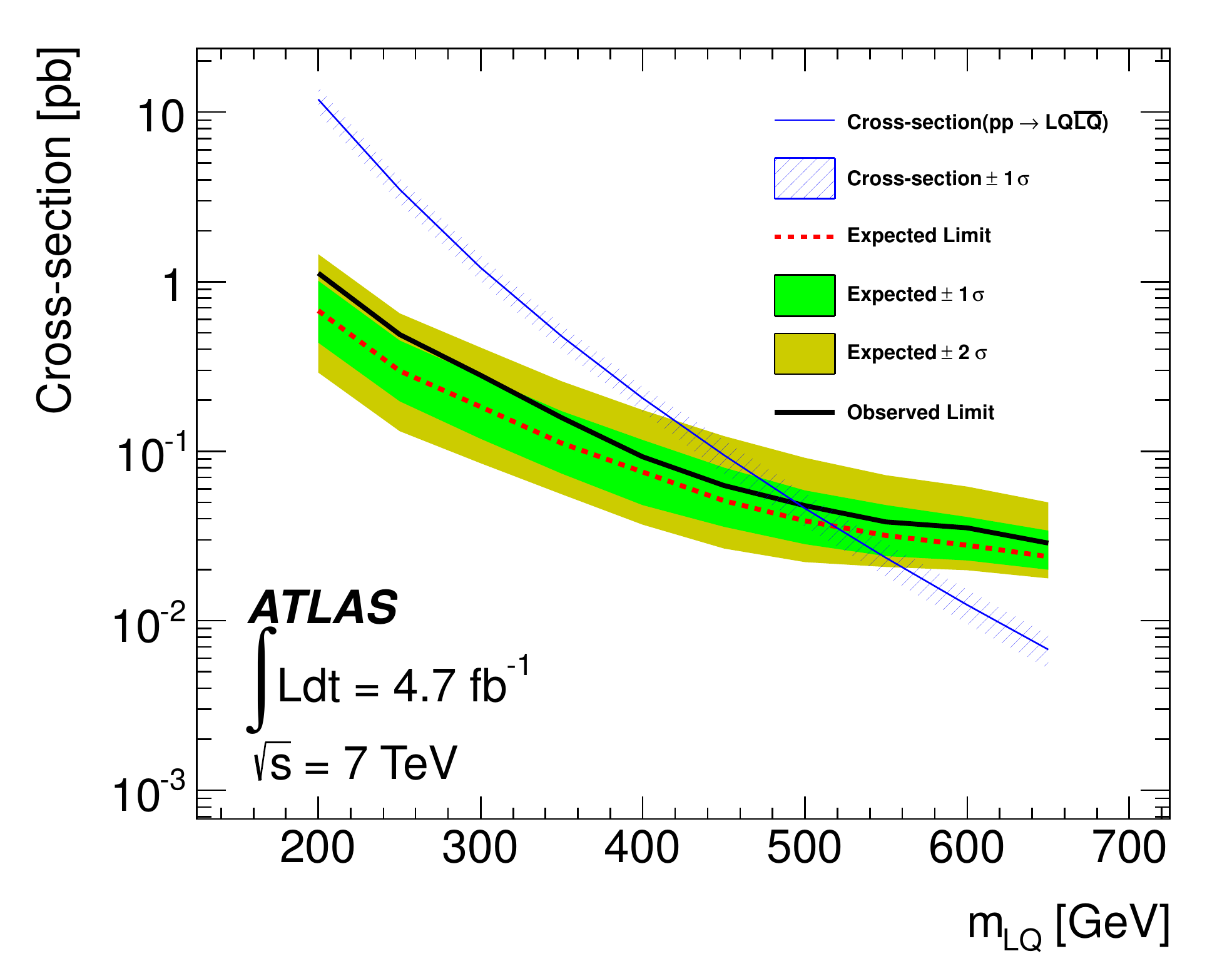}}
\subfigure[Muon Channel]{\label{fig:limit_final_muon}\includegraphics[width=0.49\textwidth]{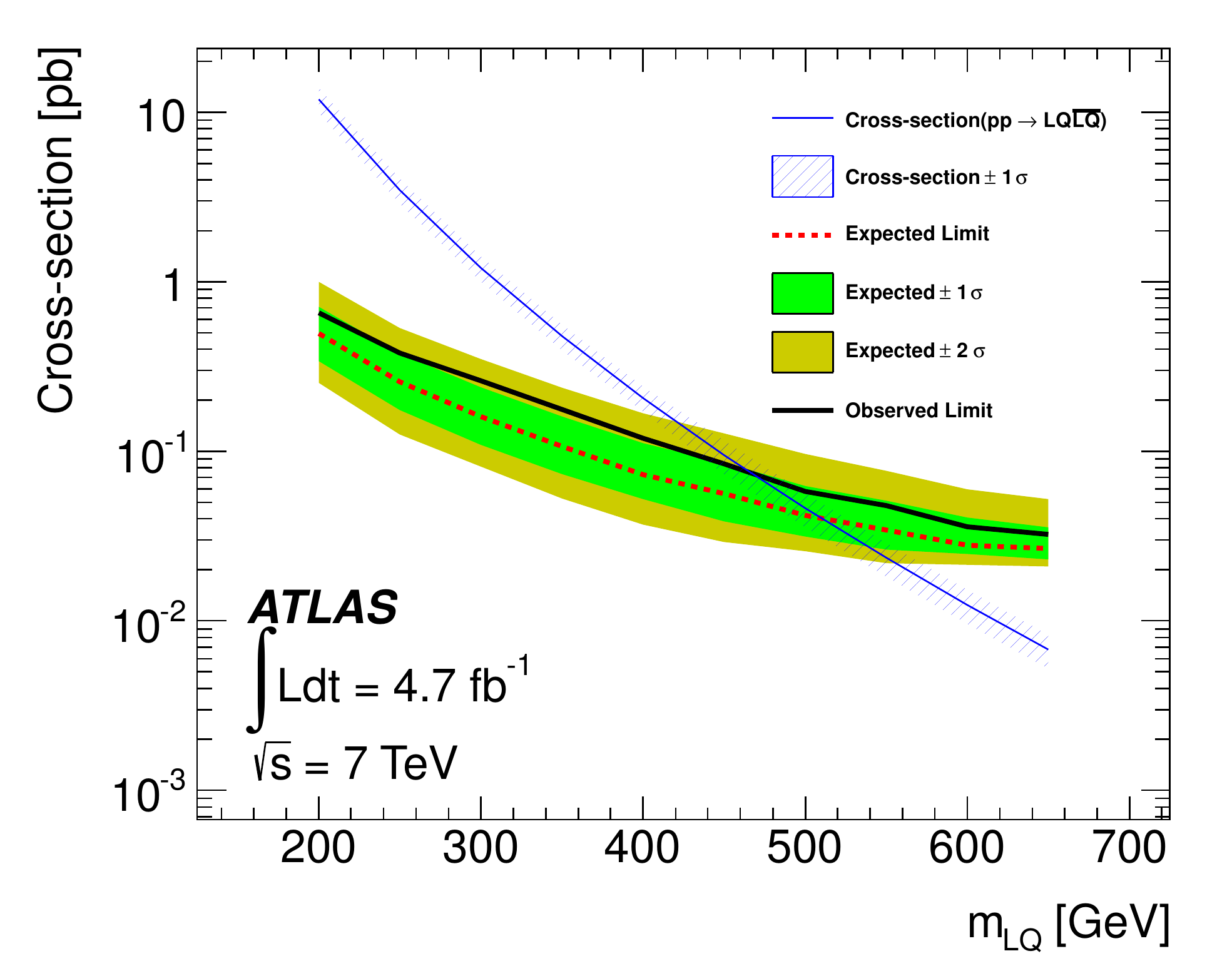}}
\caption{The expected (dashed) and observed (solid) 95\% credibility upper limits on the cross-section as a function of leptoquark mass, in the \subref{fig:limit_final_elec} electron and \subref{fig:limit_final_muon} muon channels. The 1(2)~$\sigma$~error bands on the expected limit represent all sources of systematic and statistical uncertainty.  The expected NLO production cross-section for third generation scalar leptoquarks and its corresponding theoretical uncertainty (hashed band) are also included.}
\label{fig:limit}
\end{figure}

Two  alternative models  are built to  describe background-only  and
signal+background   hypotheses.   The   signal   component  is   calculated
separately for each leptoquark  mass, thus taking the mass dependence of
the \ST~distribution into  account.  For each mass hypothesis, a single `signal strength'  parameter ($\mu$) multiplies the expected signal in each bin, where $\mu$
= 0 corresponds to the absence  of a signal and $\mu$ = 1 corresponds to
the presence of a signal with  nominal strength. The model describes the
expected  number of signal  ($s_i$) and  background ($b_i$)  events in
each bin using a Poisson  distribution.  All systematic  uncertainties described in section \ref{sec:syst}~are
assumed  to  be  distributed  according  to a  Gaussian  function  and
implemented as multiplicative constraint terms. The correlation of the
systematic uncertainties  across channels is taken  into account.  Statistical uncertainties in signal and background histogram bins are also  treated as  nuisance parameters  and assumed  to  be distributed
according to a Poisson function.  The statistical analysis of the data
employs  a  binned   likelihood  function  $\mathcal{L}(\mu,  \theta)$.
The likelihood  in each  channel is  a product over  bins in  the \ST~
distributions defined as

\begin{equation}
\mathcal {L}(\mu, \theta) =\prod_{i=\rm{bin}}{\rm{Poisson}}(N_i|\mu s_i + b_i),
\end{equation}

\noindent where $s_i$ and $b_i$ are the expected number of signal and background events in bin $i$ respectively, and $N_i$ is the observed number of events.  Both $s_i$ and $b_i$ depend on nuisance parameters $\boldsymbol{\theta}$.  Pseudo-experiments are generated according to background-only and signal+background models to obtain distributions of the test statistic, $\mathrm{log} (\mathcal{L}(\mu, \theta)/\mathcal{L}(0, \theta))$.  The CLs method~\cite{CLs} is used to calculate the p-values\footnote{The p-value is the probability of obtaining a test statistic at least as extreme as the one that was actually observed, assuming that the null hypothesis is true}. The signal strength parameter is varied iteratively to find the 95\% confidence level.
	
\begin{figure}
\centering 
\includegraphics[width=0.69\textwidth]{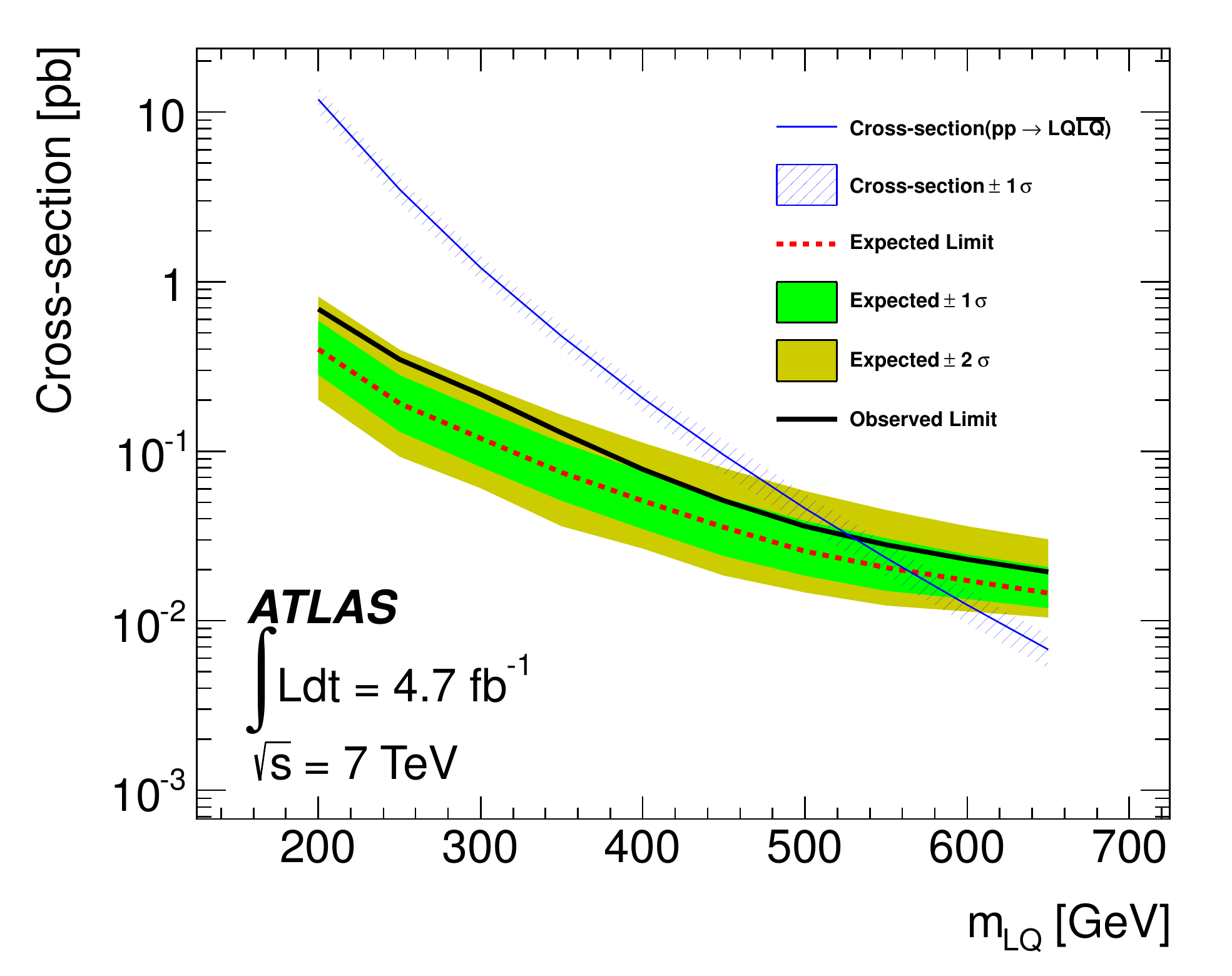}
\caption{ The expected (dashed) and observed (solid) 95\% credibility upper limits on the cross-section as a function of leptoquark mass, for the combined result.  The 1(2)~$\sigma$~error bands on the expected limit represent all sources of systematic and statistical uncertainty.  The expected NLO production cross-section for third generation scalar leptoquarks and its corresponding theoretical uncertainty (hashed band) are also included.}
\label{fig:comb}
\end{figure} 

The resulting cross-section limits as a function of leptoquark mass
are   calculated.  It is assumed that $BR (LQ\rightarrow\tau  b)$~=~1.0.  The 95\%\,CL   upper bounds  on the  NLO cross-section  for  scalar leptoquark pair production as a function of  mass are shown for individual channels in figures~\ref{fig:limit_final_elec}   and  ~\ref{fig:limit_final_muon}. Error bands for the expected limits include all sources of uncertainty.
Third   generation  scalar leptoquarks   are  observed (expected) to be excluded at 95\% confidence level for masses below 498~(523)\,\GeV~and 473~(514)\,\GeV~in the electron and  muon channels respectively by comparing the limits with theoretical predictions of cross-section versus $m_{LQ}$.  The limit is taken using the nominal theoretical calculation for the leptoquark production cross-section at NLO.  The uncertainty on the cross-section is also shown. The result when both channels are combined is shown in figure~\ref{fig:comb}. The likelihood for the combined model is defined as the product of likelihood terms for each channel.  The  data are found to be  consistent with the background-only hypothesis and third generation scalar leptoquark production is excluded at 95\%\,confidence level for leptoquark masses  up to 534\,\GeV~(the expected limit is 569\,\GeV).  

\section{Conclusions}
A search for pair production of third generation scalar leptoquarks has been performed with the ATLAS detector at the LHC, using a data sample corresponding to an integrated luminosity of 4.7\,\ifb.  No significant excess over the SM background expectation is observed in the data.  The assumption is made that $BR(LQ\rightarrow\tau  b)$~=~1.0 and third generation scalar leptoquarks with masses up to $m_{\LQ}<$~534\,\GeV~are excluded at 95$\%$\,CL.  The cross-section for leptoquark pair-production increases with centre-of-mass collision energy.  At $\sqrt{s}$~=~8\,\TeV~the production rate for leptoquarks with $m_{LQ}$~=~700\,\GeV~is enhanced by a factor of two, providing scope for setting stronger limits using data from the 2012 LHC run.  Meanwhile, this result is the most stringent limit arising from direct searches for third generation scalar leptoquarks to-date.  

% Version 1-Apr-2013

\section{Acknowledgements}

We thank CERN for the very successful operation of the LHC, as well as the
support staff from our institutions without whom ATLAS could not be
operated efficiently.

We acknowledge the support of ANPCyT, Argentina; YerPhI, Armenia; ARC,
Australia; BMWF and FWF, Austria; ANAS, Azerbaijan; SSTC, Belarus; CNPq and FAPESP,
Brazil; NSERC, NRC and CFI, Canada; CERN; CONICYT, Chile; CAS, MOST and NSFC,
China; COLCIENCIAS, Colombia; MSMT CR, MPO CR and VSC CR, Czech Republic;
DNRF, DNSRC and Lundbeck Foundation, Denmark; EPLANET, ERC and NSRF, European Union;
IN2P3-CNRS, CEA-DSM/IRFU, France; GNSF, Georgia; BMBF, DFG, HGF, MPG and AvH
Foundation, Germany; GSRT and NSRF, Greece; ISF, MINERVA, GIF, DIP and Benoziyo Center,
Israel; INFN, Italy; MEXT and JSPS, Japan; CNRST, Morocco; FOM and NWO,
Netherlands; BRF and RCN, Norway; MNiSW, Poland; GRICES and FCT, Portugal; MERYS
(MECTS), Romania; MES of Russia and ROSATOM, Russian Federation; JINR; MSTD,
Serbia; MSSR, Slovakia; ARRS and MIZ\v{S}, Slovenia; DST/NRF, South Africa;
MICINN, Spain; SRC and Wallenberg Foundation, Sweden; SER, SNSF and Cantons of
Bern and Geneva, Switzerland; NSC, Taiwan; TAEK, Turkey; STFC, the Royal
Society and Leverhulme Trust, United Kingdom; DOE and NSF, United States of
America.

The crucial computing support from all WLCG partners is acknowledged
gratefully, in particular from CERN and the ATLAS Tier-1 facilities at
TRIUMF (Canada), NDGF (Denmark, Norway, Sweden), CC-IN2P3 (France),
KIT/GridKA (Germany), INFN-CNAF (Italy), NL-T1 (Netherlands), PIC (Spain),
ASGC (Taiwan), RAL (UK) and BNL (USA) and in the Tier-2 facilities
worldwide.

\onecolumn
\clearpage 
% ATLAS Collaboration author list
% Data extracted on 30-Nov-2012 for paper reference EXOT-2012-14
%\documentclass[11pt]{article}
%\usepackage{a4wide}\begin{document}
\begin{flushleft}
{\Large The ATLAS Collaboration}

\bigskip

G.~Aad$^{\rm 48}$,
T.~Abajyan$^{\rm 21}$,
B.~Abbott$^{\rm 111}$,
J.~Abdallah$^{\rm 12}$,
S.~Abdel~Khalek$^{\rm 115}$,
A.A.~Abdelalim$^{\rm 49}$,
O.~Abdinov$^{\rm 11}$,
R.~Aben$^{\rm 105}$,
B.~Abi$^{\rm 112}$,
M.~Abolins$^{\rm 88}$
% ...
\bigskip
\\
% ...
$^{11}$ Institute of Physics, Azerbaijan Academy of Sciences, Baku, Azerbaijan\\
$^{12}$ Institut de F{\'\i}sica d'Altes Energies and Departament de F{\'\i}sica de la Universitat Aut{\`o}noma de Barcelona and ICREA, Barcelona, Spain\\
% ...
$^{21}$ Physikalisches Institut, University of Bonn, Bonn, Germany\\
% ...
$^{48}$ Fakult{\"a}t f{\"u}r Mathematik und Physik, Albert-Ludwigs-Universit{\"a}t, Freiburg, Germany\\
$^{49}$ Section de Physique, Universit{\'e} de Gen{\`e}ve, Geneva, Switzerland\\
% ...
$^{88}$ Department of Physics and Astronomy, Michigan State University, East Lansing MI, United States of America\\
% ...
$^{105}$ Nikhef National Institute for Subatomic Physics and University of Amsterdam, Amsterdam, Netherlands\\
% ...
$^{111}$ Homer L. Dodge Department of Physics and Astronomy, University of Oklahoma, Norman OK, United States of America\\
$^{112}$ Department of Physics, Oklahoma State University, Stillwater OK, United States of America\\
% ...
$^{115}$ LAL, Universit{\'e} Paris-Sud and CNRS/IN2P3, Orsay, France\\
% ...
\end{flushleft}

%\end{document}
% Created with ./xml2latex.py 

\end{document}